\documentclass[english,12pt]{article}
\usepackage[T1]{fontenc}
\usepackage[latin9]{inputenc}
\usepackage{geometry}
\usepackage{amsmath}
\usepackage{amsthm}
\usepackage{amssymb}
\usepackage{setspace}
\usepackage[authoryear]{natbib}
\makeatletter
\theoremstyle{plain}
\newtheorem{assumption}{\protect\assumptionname}
\theoremstyle{definition}
\newtheorem{defn}{\protect\definitionname}
\theoremstyle{plain}
\newtheorem{thm}{\protect\theoremname}
\theoremstyle{plain}
\newtheorem{prop}{\protect\propositionname}

\DeclareMathOperator{\E}{E}

\DeclareMathOperator{\MTE}{MTE}

\DeclareMathOperator{\ATE}{ATE}

\DeclareMathOperator{\LATE}{LATE}
\DeclareMathOperator{\LATT}{LATT}
\DeclareMathOperator{\LATUT}{LATUT}
\DeclareMathOperator{\LIV}{LIV}

\usepackage{xcolor}
\definecolor{DodgerBlue4}{RGB}{16,78,139}
\usepackage{hyperref}
\hypersetup{
	linktoc = all,
	colorlinks = true,
	urlcolor = DodgerBlue4,
	citecolor = DodgerBlue4,
	linkcolor = DodgerBlue4
}

\usepackage{booktabs}
\usepackage{siunitx}
\newcolumntype{d}{S[
    input-open-uncertainty=,
    input-close-uncertainty=,
    parse-numbers = false,
    table-align-text-pre=false,
    table-align-text-post=false
 ]}

\makeatother

\usepackage{babel}
\providecommand{\assumptionname}{Assumption}
\providecommand{\definitionname}{Definition}
\providecommand{\propositionname}{Proposition}
\providecommand{\theoremname}{Theorem}

\begin{document}
\title{Marginal Treatment Effects and Monotonicity}
\author{Henrik Sigstad\thanks{BI Norwegian Business School, Department of Economics (e-mail: henrik.sigstad@bi.no).
Thanks to Magne Mogstad and Vitor Possebom.}}
\maketitle
\begin{abstract}
How robust are analyses based on marginal treatment effects (MTE)
to violations of \citet{Imbens1994Identification} monotonicity? In
this note, I present weaker forms of monotonicity under which popular
MTE-based estimands still identify the parameters of interest.
\end{abstract}

\section{Introduction}

Marginal treatment effects (MTE), introduced by \citet{bjorklund1987estimation}
and generalized by \citet{heckman1999local,heckman2005structural},
provide a unified way of estimating various treatment effects with
continuous instruments. For instance, MTE analysis can be used to
identify the average treatment effect, the average treatment effect
on the treated and the untreated, and other policy-relevant treatment
effects. In contrast, with a continuous instrument, two-stage least
squares identifies a convex combination of treatment effects that
is not necessarily of policy interest \citep{heckman2007econometricII}.
MTE analysis, however, relies on \citet{Imbens1994Identification}
monotonicity---often a strong assumption. For instance, in the context
of judge IV designs, \citet{Imbens1994Identification} monotonicity
requires that each judge is weakly stricter than more lenient judges
\emph{in each case}. Thus, if Judge A is stricter than Judge B in
one case, Judge A can not be more lenient than Judge B in another
case. As shown in \citet{sigstad2023monotonicity}, this assumption
is frequently violated among judges.

It is thus important to understand how MTE-based treatment effect
estimates are affected by monotonicity violations. In this note, I
derive necessary and sufficient monotonicity conditions for MTE-based
estimates of popular treatment effects to identify the parameters
of interest. Fortunately, it turns out that even when Imbens-Angrist
monotonicity is violated, MTE-based estimates of these parameters
might still be consistent. I first consider MTE-based estimates of
LATE---the average treatment effect for agents affected by the instrument.
The necessary and sufficient condition for MTE analysis to identify
LATE is that monotonicity holds between the two most extreme instrument
values. For instance, in the random-judge design, this condition requires
that the strictest judge is always stricter than the most lenient
judge. As shown in \citet{sigstad2023monotonicity}, this condition
is much more plausible in random-judge designs than \citet{Imbens1994Identification}
monotonicity. Thus, even though conventional MTE analysis assumes
Imbens-Angrist monotonicity, MTE-based LATE estimates can be highly
robust to plausible levels of monotonicity violations.

Next, I consider estimates of the average treatment effect on the
treated (ATT) and the untreated (ATUT) for the complier population.
MTE-based ATT (ATUT) estimates are consistent as long as Imbens-Angirst
monotonicity holds for all pairs of instrument values involving the
lowest (highest) instrument value. For instance, in the random-judge
design, these conditions require that the most lenient (stringent)
judge is most lenient (stringent) in all cases. These conditions are
more demanding than the condition required to estimate LATE. Estimates
of ATT and ATUT are thus more sensitive to monotonicity violations.

I also consider MTE-based estimates of the average treatment effect
(ATE), which require extrapolation beyond the observed instrument
values. As long as this extrapolation is well specified, MTE-based
ATE estimates are consistent without any monotonicity assumption.
Such estimates are equivalent to the \citet{arnold2021measuring}
approach to estimating average treatment effects. Finally, I consider
the use of MTEs to assess heterogeneous treatment effects by treatment
propensity. As long as attention is limited to aggregate properties
of the MTE curve, this practice also requires only mild monotonicity
assumptions.

While these analyses show that MTE-based estimators are relatively
robust to monotonicity violations, the intermediate step of estimating
marginal treatment effects is not a meaningful exercise when monotonicity
is violated. Instead, I propose to directly estimate the relevant
treatment parameters without first estimating an ``MTE curve''.
I show that whenever MTE analysis identifies LATE, LATE is identified
by a standard Wald estimand: the difference in average outcomes between
agents receiving the highest instrument value and agents receiving
the lowest instrument value divided by the difference in treatment
propensities. Similar results are obtained for the average treatment
effects on the treated and on the untreated for the complier population.
There are several reasons to prefer such a direct estimation of treatment
parameters over MTE-based estimation when monotonicity is violated.
First, the direct estimation is more honest and clarifies the necessary
identification assumptions. Second, the direct estimates can easily
be estimated non-parametrically and do not require a fully continuous
instrument. Finally, by targeting a specific parameter rather than
the full MTE curve, the parameter can be more precisely estimated.

\section{Marginal Treatment Effects and Monotonicity\label{sec:theory}}

Fix a probability space with the outcome corresponding to a randomly
drawn agent. Define the following random variables: A binary treatment
$D\in\left\{ 0,1\right\} $, an outcome $Y\in\mathbb{R}$, and a continuous
instrument $Z\in\mathbb{R}$ with support $\left[\underline{z},\bar{z}\right]$.
To capture the idea that different agents might be induced into treatment
in different ways by the instrument, define a \emph{response type}
as a mapping $s:\left[\underline{z},\bar{z}\right]\rightarrow\left\{ 0,1\right\} $
from instrument values to treatments.\footnote{Response types were introduced by \citet{heckman2018unordered}.}
Denote by the random variable $S$ the response type of the randomly
drawn agent. If $S=s$ for agent $i$, then $s\left(z\right)=1$ indicates
that agent $i$ will receive the treatment if $Z$ is set to $z$.
 Denote by $\mathcal{S}$ the set of all response types in the population.
Define $Y\left(0\right)$ and $Y\left(1\right)$ as the \emph{potential
outcomes} when $D$ is set to $0$ and $1$, respectively. Denote
by the random variable $\beta\equiv Y\left(1\right)-Y\left(0\right)$
the treatment effect of agent $i$. Let $p\left(z\right)\equiv\E\left[S\left(z\right)\right]$
be the share of agents receiving treatment at $Z=z$. I assume the
following throughout
\begin{assumption}
\label{assu:iv}(Exogeneity and Exclusion). $\left\{ Y\left(0\right),Y\left(1\right),S\right\} \perp Z$
\end{assumption}
\begin{assumption}
\label{assu:rank}(First stage). The propensity $p\left(z\right)$
is non-trivial function of $z$.
\end{assumption}
To simplify the notation, assume (without loss of generality) that
the instrument values are labeled such that
\begin{assumption}
\label{assu:indexed_by_stringency-1-1} $p\left(z\right)=z$.
\end{assumption}
\citet{Imbens1994Identification} monotonicity is then defined by
\begin{defn}[Imbens-Angrist Monotonicity]
 For all $z,z'\in\mathbb{R}$ and $s\in\mathcal{S}$
\[
z\geq z'\Rightarrow s\left(z\right)\geq s\left(z'\right)
\]
\end{defn}
Marginal treatment effects were introduced by \citet{bjorklund1987estimation}
and generalized by \citet{heckman1999local}.\footnote{See \citet{heckman2007econometricII}.}
In applied work (\emph{e.g.}, \citealt{Arnold2018Racial,Bhuller2020Incarceration}),
marginal treatment effect analysis relies on a generalized \citet{roy1951some}
selection model
\[
D=\mathbf{1}\left[Z>U\right]
\]
where $U$ is a random variable.\footnote{See \citet{heckman2007econometricI,heckman2007econometricII}. To
simplify the exposition, I disregard covariates.} The agent receives treatment if the instrument is above the agent's
unobserved \emph{resistance to treatment }$U$. This model implicitly
assumes Imbens-Angrist monotonicity.\footnote{Consider two instrument values $z_{1}\geq z_{2}$. Then $D\left(z_{2}\right)\geq D\left(z_{1}\right)$
for all agents.} A \emph{marginal treatment effect} is then defined as the average
treatment effect for agents with a given treatment propensity:
\[
\MTE\left(u\right)=\E\left[\beta\mid U=u\right]
\]

Marginal treatment effects can then be identified using the local
instrumental variable approach \citep{heckman1999local,heckman2005structural}:
\[
\LIV\left(u\right)\equiv\frac{d\E\left[Y\mid Z=u\right]}{du}
\]
Assume this derivative exists. Under Imbens-Angrist monotonicity,
we have $\LIV\left(u\right)=\MTE\left(u\right)$. But $\LIV\left(u\right)$
is defined even when Imbens-Angrist monotonicity does not hold.

The applied literature uses MTE analysis for two purposes. First,
to estimate meaningful treatment parameters such as LATE and ATE (\emph{e.g.},
\citealt{Arnold2018Racial,Bhuller2020Incarceration}). Second, to
assess heterogeneous treatment effects based on the treatment propensity
$U$ by directly inspecting $\LIV\left(u\right)$ (\emph{e.g.}, \citealt{doyle2007child,maestas2013does,french2014effect}).

\subsection{Using MTE to Identify Meaningful Treatment Parameters}

\citet{heckman1999local,heckman2005structural} showed that many popular
treatment parameters---including the average treatment effect (ATE)---can
be identified by a weighted average of marginal treatment effects.
MTE analysis can thus be used to identify more meaningful treatment
parameters than the weighted average of individual treatment effects
produced by 2SLS. Identifying ATEs using MTE analysis, however, requires
full support of $Z$ in $\left[0,1\right]$ or extrapolation beyond
the support of $Z$. Since $Z$ typically does not have full support
in practice, the literature instead normally seeks to estimate the
local average treatment effect (LATE) for agents receiving treatment
when $Z=\bar{z}$ and not when $Z=\underline{z}$---agents with $D\left(\underline{z}\right)<D\left(\bar{z}\right)$.
This parameter differs from the 2SLS estimand by placing equal weight
on all compliers. The literature (\emph{e.g.}, \citealt{Bhuller2020Incarceration})
has also considered the average treatment effect on the treated and
the average treatment effect on the untreated for the same population.
Since these treatment effects are ``local''---defined on the complier
population---I label them LATT and LATUT, respectively. Formally,
define

\begin{eqnarray*}
\LATE & \equiv & \E\left[\beta\mid S\left(\underline{z}\right)=0,S\left(\bar{z}\right)=1\right]\\
\LATT & \equiv & \E\left[\beta\mid S\left(\underline{z}\right)=0,D=1\right]\\
\LATUT & \equiv & \E\left[\beta\mid S\left(\bar{z}\right)=1,D=0\right]
\end{eqnarray*}

The $\LATE$ parameter is the local average treatment effect for agents
receiving treatment under the highest instrument value but not under
the lowest instrument value. The $\LATT$ and $\LATUT$ parameters
are the (local) average treatment effects on the treated and the untreated
for a similar complier population.\footnote{The $\LATT$ and $\LATUT$ complier population includes all agents
except never-takers and always-takers. The $\LATE$ complier population
also ignores, for instance, response types receiving treatment under
some intermediate instrument values but not by the highest nor the
lowest instrument value. I do not see a way to identify a local average
treatment effect that covers also such compliers.} As shown by \citet{heckman1999local,heckman2005structural}, these
parameters are identified under Imbens-Angrist monotonicity by:
\begin{eqnarray*}
\tilde{\LATE} & \equiv & \frac{1}{\bar{z}-\underline{z}}\int_{\underline{z}}^{\bar{z}}\LIV\left(u\right)du\\
\tilde{\LATT} & \equiv & \frac{1}{\E\left[Z\right]-\underline{z}}\int_{\underline{z}}^{\bar{z}}\Pr\left[Z>u\right]\LIV\left(u\right)du\\
\tilde{\LATUT} & \equiv & \frac{1}{\bar{z}-\E\left[Z\right]}\int_{\underline{z}}^{\bar{z}}\Pr\left[Z<u\right]\LIV\left(u\right)du
\end{eqnarray*}
When Imbens-Angrist monotonicity is violated, however, this method
might lead to wrong conclusions. How sensitive are these estimands
to monotonicity violations? Fortunately, it turns out that MTE analysis
might still identify $\LATE$, $\LATT$, and $\LATUT$ even when Imbens-Angrist
monotonicity is violated. 

Formally, let $\mathcal{G}$ be the set of all possible joint distributions
of $\left(Y\left(1\right),Y\left(0\right),S\right)$. To allow for
arbitrary heterogeneous effects, we do not want to place any restrictions
on this joint distribution. The necessary and sufficient conditions
for $\tilde{\LATE}$, $\tilde{\LATT}$, $\tilde{\LATUT}$ to be consistent
under arbitrary heterogeneous effects are the following much weaker
monotonicity conditions:
\begin{thm}
\label{thm:id1} (Identification results).
\end{thm}
\begin{description}
\item [{i)}] $\tilde{\LATE}=\LATE$ for all $g\in\mathcal{G}$ if and only
if $s\left(\bar{z}\right)\geq s\left(\underline{z}\right)$ for all
$s\in\mathcal{S}$.
\item [{ii)}] $\tilde{\LATT}=\LATT$ for all $g\in\mathcal{G}$ if and
only if $s\left(z\right)\geq s\left(\underline{z}\right)$ for all
$s\in\mathcal{S}$ and $z$.
\item [{iii)}] $\tilde{\LATUT}=\LATUT$ for all $g\in\mathcal{G}$ if and
only if $s\left(\bar{z}\right)\geq s\left(z\right)$ for all $s\in\mathcal{S}$
and $z$.
\end{description}
Thus, for MTE analysis to identify $\LATE$, it is sufficient that
monotonicity holds between the lowest and the highest instrument value.
This condition is substantially weaker than Imbens-Angrist monotonicity,
especially when there are many possible instrument values. Similarly,
$\LATT$ is identified by MTE analysis whenever monotonicity holds
for all instrument value pairs that involve the lowest instrument
value. This condition is stronger than $\LATE$ condition but still
considerably weaker than Imbens-Angrist monotonicity---monotonicity
is allowed to be violated for all instrument value pairs that do not
include the lowest instrument value. For MTE analysis to identify
$\LATUT$, on the other hand, monotonicity must hold for all instrument
value pairs that involve the \emph{highest} instrument value.

\subsection{Estimating ATE by Extrapolating the MTE curve}

When $f\left(u\right)\equiv\E\left[Y\mid Z=u\right]$ is estimated
parametrically, one might seek to extrapolate beyond the support of
$Z$ to estimate the average treatment effect, $\ATE\equiv\E\left[\beta\right]$.
In particular, let $\hat{f}:\left[0,1\right]\rightarrow\mathbb{R}$
be an extrapolation of $f$ that covers the full interval $\left[0,1\right]$.
The corresponding MTE curve is defined by $\hat{\LIV}\left(u\right)\equiv\hat{f}'\left(u\right)$.
One can then estimate $\ATE$ by
\[
\tilde{\ATE}\equiv\int_{0}^{1}\hat{\LIV}\left(u\right)du
\]
How do monotonicity violations influence such analysis? By the fundamental
theorem of calculus, $\tilde{\ATE}=\hat{f}\left(1\right)-\hat{f}\left(0\right)$.
If the extrapolation is well specified, $\hat{f}\left(1\right)$ can
be thought of as the average outcome for agents in the hypothetical
case of receiving $Z=1$.\footnote{In the context of the judge IV design, it would correspond to the
average outcomes for defendants randomly assigned a hypothetical supremely
stringent judge that always incarcerates.} In that case, $\hat{f}\left(1\right)=\E\left[Y\left(1\right)\right]$.
Similarly, $\hat{f}\left(0\right)$ can be thought of as the average
outcome for agents had they been assigned $Z=0$ which gives $\hat{f}\left(0\right)=\E\left[Y\left(0\right)\right]$.
We thus get that $\tilde{\ATE}=\E\left[Y\left(1\right)-Y\left(0\right)\right]=\ATE$
if the extrapolation is well specified---$\hat{f}$ is able to identify
the average outcome for agents in the hypothetical cases of receiving
$Z=1$ and $Z=0$. Formally
\begin{prop}
$\tilde{\ATE}=\ATE$ if $\hat{f}\left(1\right)=\E\left[Y\left(1\right)\right]$
and $\hat{f}\left(0\right)=\E\left[Y\left(0\right)\right]$.
\end{prop}
The MTE-based estimator of $\ATE$ is equivalent to the estimator
proposed by \citet{arnold2021measuring}, who extrapolate towards
a supremely lenient judge to estimate the ATE of pre-trial release
on pre-trial misconduct in a judge IV setting. As pointed out by \citet{arnold2021measuring},
this approach does not require any monotonicity assumptions. Thus,
monotonicity violations do not affect the validity of this approach.

\subsection{Using MTE to Analyze Heterogeneous Effects}

The literature also uses the MTE framework to assess heterogeneous
treatment effects based on the treatment propensity $U$ by directly
inspecting $\LIV\left(u\right)$---the ``MTE curve'' (\emph{e.g.},
\citealt{doyle2007child,maestas2013does,french2014effect}). But $\LIV\left(u\right)$
is difficult to interpret when Imbens-Angrist monotonicity is violated.
To see this, it is instructive to consider $\LIV\left(u\right)$ as
the limit of a standard Wald estimand:
\[
\LIV\left(u\right)=\lim_{v\rightarrow u}\tilde{\LATE}_{u,v}
\]
where

\[
\tilde{\LATE}_{u,v}\equiv\frac{\E\left[Y\mid Z=u\right]-\E\left[Y\mid Z=v\right]}{u-v}
\]

Under Imbens-Angrist monotonicity, $\tilde{\LATE}_{u,v}$ identifies
\[
\LATE_{u,v}\equiv\E\left[\beta\mid D\left(u\right)>D\left(v\right)\right]
\]
the local average treatment effect for cases where receiving treatment
at $Z=u$ but not at $Z=v$. As $v$ approaches $u$, however, Imbens-Angrist
monotonicity between $v$ and $u$ might be unlikely.\footnote{In the context of judges, Imbens-Angrist monotonicity is less likely
for judge pairs with similar stringencies than for judge pairs with
more different stringencies \citep{sigstad2023monotonicity}. } Individual points of the MTE curve are then hard to interpret. But
looking at more aggregate properties of the MTE curve could still
be meaningful. For instance, the average of $\LIV\left(u\right)$
across a range $u\in\left[\underline{u},\bar{u}\right]$ identifies
LATE for agents receiving treatment at $Z=\bar{u}$ but not at $Z=\underline{u}$
when monotonicity holds between these instrument values:
\begin{prop}
\label{prop:MTE-curve}
\[
\E\left[\LIV\left(U\right)\mid\underline{u}\leq U\leq\bar{u}\right]=\LATE_{\underline{u},\bar{u}}
\]
for all $g\in\mathcal{G}$ if and only if $s\left(\bar{u}\right)\geq s\left(\underline{u}\right)$
for all $s\in\mathcal{S}$.
\end{prop}

As $\bar{u}$ and $\underline{u}$ become more distant, monotonicity
between these two values typically becomes more plausible.\footnote{This is at least true for the random-judge design \citep{sigstad2023monotonicity}.}

\subsection{Identifying Meaningful Treatment Effects without MTE\label{subsec:without-MTE-1}}

While MTE analysis gives correct results under weaker assumptions
than Imbens-Angrist monotonicity, the ``MTE curve'' $\LIV\left(u\right)$
is not a meaningful object when monotonicity is violated. A more honest
approach is to estimate aggregate treatment effects directly, without
first estimating an MTE curve. The following results show how LATE,
LATT, and LATUT can be directly identified without first estimating
$\LIV\left(u\right)$.
\begin{thm}
\label{thm:id2} (Identifying meaningful treatment effects without
MTE analysis).
\begin{description}
\item [{i)}] $\LATE=\frac{\E\left[Y\mid Z=\bar{z}\right]-\E\left[Y\mid Z=\underline{z}\right]}{\bar{z}-\underline{z}}$
if $s\left(\bar{z}\right)\geq s\left(\underline{z}\right)$ for all
$s\in\mathcal{S}$.
\item [{ii)}] $\LATT=\frac{\E\left[Y\right]-\E\left[Y\mid Z=\underline{z}\right]}{\E\left[Z\right]-\underline{z}}$
if $s\left(z\right)\geq s\left(\underline{z}\right)$ for all $s\in\mathcal{S}$
and $z$.
\item [{iii)}] $\LATUT=\frac{\E\left[Y\mid Z=\bar{z}\right]-\E\left[Y\right]}{\bar{z}-\E\left[Z\right]}$
if $s\left(\bar{z}\right)\geq s\left(z\right)$ for all $s\in\mathcal{S}$
and $z$.
\item [{iv)}] $\LATE_{z_{1},z_{2}}=\frac{\E\left[Y\mid Z=z_{1}\right]-\E\left[Y\mid Z=z_{2}\right]}{z_{1}-z_{2}}$
if $s\left(z_{1}\right)\geq s\left(z_{2}\right)$ for all $s\in\mathcal{S}$.
\end{description}
\end{thm}
In particular, $\LATE$ is identified by the standard Wald estimand
of the effect of receiving the highest instrument value compared to
receiving the lowest instrument value.\footnote{A similar estimand is discussed by \citet{frolich2007nonparametric}
(Theorem 8).} Furthermore, $\LATT$ and $\LATUT$ are identified by the difference
between the mean outcome and the expected outcomes for agents receiving
the lowest and highest instrument values, respectively. The only parameters
that need to be estimated are thus $\bar{z}$, $\underline{z}$, $\E\left[Y\mid Z=\underline{z}\right]$
and $\E\left[Y\mid Z=\bar{z}\right]$. There are two advantages of
this approach. First, it is not needed to estimate a full MTE curve.
Estimating an MTE curve is difficult in practice due to data limitations
and typically requires parametric assumptions. When the aim is to
estimate only $\E\left[Y\mid Z=\underline{z}\right]$ and $\E\left[Y\mid Z=\bar{z}\right]$
instead of the full MTE curve, one can do this non-parametrically.\footnote{For instance, one can directly estimate $\E\left[Y\mid Z=\bar{z}\right]$
using the sample analog, or one can estimate it using a local linear
regression (with, \emph{e.g.}, a triangular kernel) on a sample of
the highest instrument values. Note that such LATE estimates (obtained
either through MTE analysis or the Wald approach) essentially ignore
agents receiving medium instrument values. These estimates will thus
typically be less precise than 2SLS estimates which exploit all instrument
values. Also, note that in finite samples, the sample analog of $\bar{z}-\underline{z}$
will be upward biased. For instance, even if all instrument values
have the same treatment propensity ($\bar{z}=\underline{z}$), the
sample analog of $\bar{z}-\underline{z}$ will still be positive.
To avoid this bias, one could use a split-sample approach: Estimate
which instrument values are associated with the highest and lowest
treatment propensities in one sample and estimate the treatment propensities
associated with these instrument values in another sample. I leave
a thorough discussion of inference to future research.} Second, the results above are valid also for discrete instruments
when MTE analysis is not applicable.\footnote{For instance, applying MTE-analysis in the judge IV setting formally
requires a continuum of judges.}

\section{Conclusions\label{sec:conclusions}}

Marginal treatment effects can be used to estimate more meaningful
treatment parameters than two-stage least squares but require Imbens-Angrist
monotonicity. In this note, I have presented conditions under which
MTE-based estimates still identify the parameters of interest when
Imbens-Angrist monotonicity is violated. I also showed how the same
parameters can be identified without relying on the MTE framework.
I leave questions of estimation for future work.

\small

\bibliographystyle{authordate3}
\bibliography{references}

\normalsize

\clearpage{}

\appendix

\section{Proofs}
\begin{proof}
(Theorem \ref{thm:id1}.) \emph{Part i).}
\begin{eqnarray*}
\tilde{\LATE} & = & \frac{1}{\bar{z}-\underline{z}}\int_{\underline{z}}^{\bar{z}}\frac{d\E\left[Y\mid Z=u\right]}{du}du\\
 & = & \frac{1}{\bar{z}-\underline{z}}\left(\E\left[Y\mid Z=\bar{z}\right]-\E\left[Y\mid Z=\underline{z}\right]\right)\\
 & = & \frac{1}{\bar{z}-\underline{z}}\left(\E\left[DY\left(1\right)+\left(1-D\right)Y\left(0\right)\mid Z=\bar{z}\right]-\E\left[DY\left(1\right)-\left(1-D\right)Y\left(0\right)\mid Z=\underline{z}\right]\right)\\
 & = & \frac{\Pr\left[D\left(\bar{z}\right)>D\left(\underline{z}\right)\right]}{\bar{z}-\underline{z}}\E\left[Y\left(1\right)-Y\left(0\right)\mid D\left(\bar{z}\right)>D\left(\underline{z}\right)\right]\\
 & - & \frac{\Pr\left[D\left(\bar{z}\right)<D\left(\underline{z}\right)\right]}{\bar{z}-\underline{z}}\E\left[Y\left(1\right)-Y\left(0\right)\mid D\left(\bar{z}\right)<D\left(\underline{z}\right)\right]
\end{eqnarray*}
The first equality invokes the fundamental theorem of calculus and
the fourth equality uses Assumption \ref{assu:iv}. This expression
equals $\LATE$ for all $g\in\mathcal{G}$ if and only if $\Pr\left[D\left(\bar{z}\right)<D\left(\underline{z}\right)\right]=0$.

\emph{Part ii).} \emph{Let $f\left(u\right)$ denote the density of
$Z$ at $u$. Then}
\begin{eqnarray*}
\tilde{\LATT} & = & \frac{1}{\E\left[Z\right]-\underline{z}}\int_{\underline{z}}^{\bar{z}}\Pr\left[Z>u\right]\frac{d\E\left[Y\mid Z=u\right]}{du}du\\
 & = & \frac{1}{\E\left[Z\right]-\underline{z}}\int_{\underline{z}}^{\bar{z}}\left(\E\left[Y\mid Z=u\right]-\E\left[Y\mid Z=\underline{z}\right]\right)f\left(u\right)du\\
 & = & \frac{1}{\E\left[Z\right]-\underline{z}}\left(\E\left[Y\right]-\E\left[Y\mid Z=\underline{z}\right]\right)\\
 & = & \frac{1}{\E\left[Z\right]-\underline{z}}\E\left[\E\left[Y\mid S\right]-\E\left[Y\mid Z=\underline{z},S\right]\right]\\
 & = & \frac{1}{\E\left[Z\right]-\underline{z}}\E\left[\E\left[DY\left(1\right)+\left(1-D\right)Y\left(0\right)\mid S\right]-\E\left[Y\mid Z=\underline{z},S\right]\right]\\
 & = & \frac{1}{\E\left[Z\right]-\underline{z}}\E\left[\E\left[D\mid S\right]\E\left[Y\left(1\right)\mid S\right]+\left(1-\E\left[D\mid S\right]\right)\E\left[Y\left(0\right)\mid S\right]-\E\left[Y\mid Z=\underline{z},S\right]\right]\\
 & = & \frac{1}{\E\left[Z\right]-\underline{z}}\E\left[\E\left[D\mid S\right]\E\left[Y\left(1\right)-Y\left(0\right)\mid S\right]+\E\left[Y\left(0\right)\mid S\right]-\E\left[Y\mid Z=\underline{z},S\right]\right]
\end{eqnarray*}
The second equality uses that both integrals represent the area between
the curve $\E\left[Y\mid Z=u\right]$ and $\E\left[Y\mid Z=\underline{z}\right]$
(weighted by the density $f$). The fourth equality uses the law of
iterated expectations, and the sixth equality invokes Assumption \ref{assu:iv}.
For this to equal 
\begin{eqnarray*}
\LATT & = & \E\left[Y\left(1\right)-Y\left(0\right)\mid D\left(\underline{z}\right)<D\left(\bar{z}\right),D=1\right]\\
 & = & \frac{1}{\E\left[D\mid D\left(\underline{z}\right)<D\left(\bar{z}\right)\right]}\E\left[D\left(Y\left(1\right)-Y\left(0\right)\right)\mid D\left(\underline{z}\right)<D\left(\bar{z}\right)\right]
\end{eqnarray*}
 for all $g\in\mathcal{G}$, we need that for each $s\in\mathcal{S}$,
either $s\left(\underline{z}\right)=0$ or $\E\left[D\mid S=s\right]=1$.\footnote{If $\E\left[D\mid S=s\right]<1$ and $s\left(\underline{z}\right)=1$
for a response type $s\in\mathcal{S}$, $\tilde{\LATT}$ will---unlike
$\LATT$---place a negative weight on $\E\left[Y\left(1\right)-Y\left(0\right)\mid S=s\right]$.
It is straightforward to check that the expressions for $\tilde{\LATT}$
and $\LATT$ coincide when either $s\left(\underline{z}\right)=0$
or $\E\left[D\mid S=s\right]=1$ for all $s\in\mathcal{S}$.} In other words, we need $D\left(z\right)\geq D\left(\underline{z}\right)$
for all $z$. The proof of part iii) is analogous to part ii).
\end{proof}
\begin{proof}
(Theorem \ref{thm:id2}.) This follows from the proof of Theorem \ref{thm:id1}.
\end{proof}
\begin{proof}
(Proposition \ref{prop:MTE-curve}.)
\begin{eqnarray*}
\E\left[\LIV\left(U\right)\mid\underline{u}\leq U\leq\bar{u}\right] & = & \frac{1}{\bar{u}-\underline{u}}\int_{\underline{u}}^{\bar{u}}\LIV\left(u\right)du\\
 & = & \frac{\E\left[Y\mid Z=\bar{u}\right]-\E\left[Y\mid Z=\underline{u}\right]}{\bar{u}-\underline{u}}
\end{eqnarray*}

The latter Wald estimand identifies $\LATE_{\underline{u},\bar{u}}$
for all $g\in\mathcal{G}$ if and only if there are no ``defiers'',
$\Pr\left[D\left(\underline{u}\right)>D\left(\bar{u}\right)\right]=0$.
\end{proof}

\end{document}